\documentclass{raa_twocolumn}            

\usepackage{graphicx,amsmath,amssymb}             

\begin{document}

   \title{Pulsar radio emission height as a function of phase
}

   \volnopage{Vol.0 (20xx) No.0, 000--000}      
   \setcounter{page}{1}          

   \author{H. Tong, H. G. Wang}

   \institute{School of Physics and Materials Science, Guangzhou University, Guangzhou 510006, China;
   {\it tonghao@gzhu.edu.cn}\\
   }

   \date{Received~~2009 month day; accepted~~2009~~month day}

\abstract{We present an analytical treatment of the pulsar radio emission height as a function of phase based on a set of simplifying assumptions. Assuming a dipole field geometry, the emission height can be expressed as a function of phase and the impact angle. We found that: (1) The emission height is a quadratic function of the phase, given the magnetospheric geometry. The emission height is higher at the edge of the pulse profile than at the center. (2) The emission height is also a quadratic function of the impact angle. This point can be compared directly with the observations of the geodetic precessing pulsar PSR J1906+0746. (3) Larger inclination angle may imply a higher emission height, when other parameters are similar. (4) By assuming curvature radiation, or inverse Compton scattering, the Lorentz factor of the radio emitting particles can be obtained as a function of phase.
}

   \authorrunning{Tong \& Wang}            
   \titlerunning{Pulsar radio emission height as a function of phase}  

   \maketitle

%
%
\section{Introduction}
The magnetospheric structure and radio emission mechanism is a long standing problem of pulsar studies (Philippov \& Kramer 2022). This problem is renewed by the observations of fast radio bursts, which may share similar magnetospheric physics of pulsars (Zhang 2023). From the observational point of view, radio observations have tried to determine the three-dimensional structure of the emission region. The rotating vector model is employed for the polarization position angle of pulsars (Radhakrishan \& Cooke 1969; Johnston et al. 2023), from which the magnetospheric geometry may be determined, mainly the inclination angle $\alpha$ (angle between the rotational axis and magnetic axis) and impact angle $\beta$ (closest approach between the magnetic axis and line of sight). The radius-to-frequency mapping is proposed for the radio emission height
as a function of frequency (Ruderman \& Sutherland 1979; Cordes 1978; Wang et al. 2013; Qiu et al. 2023). Various beam forms are envisaged (e.g., core+cone, patchy, fan beam etc, Rankin 1993; Lyne \& Manchester 1988; Wang et al. 2014).

In order to obtain the three dimensional information of the emission region, the emission height should be determined. It can be determined under different assumptions. The emission height may be obtained by using the pulse width information (Mitra \& Rankin 2002). Considering a rotating magnetosphere, the emission height may be determined by measuring the phase shift between polarization position angle curve and intensity curve (Blaskiewicz et al. 1991; Johnston \& Kramer 2019). A finite emission height will result in aberration (a rotating magnetosphere) and retardation. The emission height may be determined by the relativistic phase shift of the leading and trailing component of the emission cones (Gangadhara \& Gupta 2001; Dyks et al. 2004; Roy et al. 2025).

Recent observational progresses have provided us with more information of the emission height. The relativistic phase shift method can determine the emission height for different cones at different phase (Gangadhara \& Gupta 2001; Dyks et al. 2004). The geodetic precessing pulsar PSR J1906+0746 exhibits the two-dimensional structure of the emission beam (Desvignes et al. 2019). At the same time, it also demonstrates different polarization details and emission heights for different impact angles. The emission height as a function of $\beta$ is determined in PSR J1906+0746 (Desvignes et al. 2019). With the advancement of FAST, high signal-to-noise ratio observations for more pulsars are available. In order to compare with the FAST observations, the emission height and Lorentz factor as a function of pulse phase are calculated numerically (Wang et al. 2024; Cao et al. 2024).

In principle, previous theoretical modeling of pulsar emission height already contain the information of emission height as a function of phase and $\beta$ etc (Gangadhara \& Gupta 2001; Dyks et al. 2004; Gangadhara 2004; Wang et al. 2006; Zhang et al. 2007; Lee et al. 2009; Wang et al. 2013; Yuen \& Melrose 2014; Roy 2023). But in these previous treatments, either only one emission height is given (e.g., at the edge of the pulse profile, or at the core of pulse profile), or the modeling tries to be as rigorous as possible, which hinders a direct and quick comparison with observations. We note that the emission beam radius, the pulse profile width and the impact angle, etc, are always small quantities. Therefore, based on previous works, by performing Taylor expansion of these small parameters, an analytical expression for the emission height as a function of phase and $\beta$ is obtained.

\section{Pulsar radio emission height as a function of phase}

\subsection{Analytical treatment}
Our treatment is based on Zhang et al. (2007). Similar treatment are also employed by the observers to compare it with observations (Lee et al. 2009; Lu et al. 2016; Shang et al. 2021; Wang et al. 2024; Cao et al. 2024). Different from these previous works, we try to provide a simplified and analytical treatment.

Assuming a dipole field geometry, the emission height is known once the field line and the magnetic colatitude is known:
\begin{equation}
  r = R_e \sin^2\theta,
\end{equation}
where $R_e$ is the maximum radial extension of the field line, $\theta$ is the magnetic colatitude of the emission point (in the magnetic frame).
Observationally, the duty cycle of the pulsar radio emission is relative small (e.g., $< 0.1$, Johnston \& Kramer 2019). This implies that the radio emission point may be near the magnetic axis. Therefore, the magnetic colatitude $\theta$ is a small quantity. For small $\theta$, the relation between the emission point and line of sight is (Hibschman \& Arons 2001; Tong et al. 2021):
\begin{equation}\label{eqn_theta_thetaobs_relation}
  \theta = \frac23 \theta_{\rm obs},
\end{equation}
where $\theta_{\rm obs}$ is the magnetic colatitude of the line of sight. The pulsar geometry is mainly determined by the inclination angle $\alpha$ and viewing angle $\zeta\equiv \alpha+\beta$ (angle between the rotational axis and line of sight). The magnetic colatitude of the line of sight $\theta_{\rm obs}$ is related to the pulse phase as:
\begin{flalign}
  &\ \cos\theta_{\rm obs} = \cos\alpha \cos\zeta + \sin\alpha \sin\zeta \cos\phi &\\
  &\ = \scriptsize{\cos\alpha \cos(\alpha + \beta) + \sin\alpha \sin(\alpha + \beta) \cos\phi}.&
\end{flalign}
The corresponding geometry can be seen in figure 3 in Zhang et al. (2007) or figure 1 in Tong et al. (2021).
Here $\phi=0$ corresponds to the meridian plane defined by the rotational axis and magnetic axis.
Again, $\beta$ and $\phi$ may be small quantities. Then the above equation can be simplified using serial expansion:
\begin{equation}\label{eqn_thetaobs}
  \theta_{\rm obs}^2 = \sin^2\alpha \ \phi^2 + \beta^2.
\end{equation}
This is just a right triangle with three sides: $\beta$, $\sin\alpha \ \phi$ and $\theta_{\rm obs}$.
Therefore, the emission height as a function of phase is:
\begin{eqnarray}
\label{eqn_emission_height_thetaobs}
  r &=& \frac49 R_e \theta_{\rm obs}^2\\
  \label{eqn_emission_height}
  &=&\frac49 R_e (\sin^2\alpha \ \phi^2 + \beta^2).
\end{eqnarray}

The maximum radial extent of the field line $R_e$ may dependent on the inclination angle and phase (Zhang et al. 2007; Lee et al. 2009; Lu et al. 2016). But since the emission height is proportional to $\phi^2$ and $\beta^2$, then any dependence of $R_e$ on $\phi$ will be higher order modifications. Following previous treatment (Wang et al. 2006; Wang et al. 2013), we denote $R_e$ as:
\begin{equation}\label{eqn_Re}
  R_e = \frac{1}{\eta^2} R_{\rm lc},
\end{equation}
where $R_{\rm lc}$ is the light cylinder radius, $\eta$ is a dimensionless quantity. For $\eta=1$, $R_e$ corresponds to the last open field line. For $\eta=0$, $R_e$ corresponds to the magnetic axis. For $\eta=0.5$, $R_e$ corresponds to the field line with foot position at the neutron star surface $0.5$ times the polar cap radius. Other values of $\eta$ are similar. At the same time, the dimensionless parameter $\eta$ absorbs all our ignorance and approximations. For an aligned rotator, the critical field line intersects the polar cap at $\eta= 0.74$ (Rutherland \& Sutherland 1975). Subsequently, field lines with $\eta>0.74$ are named as annular region. While field lines with $\eta<0.74$ are named as core region (Qiao et al. 2004). Such a distinction is also employed by the observers (Lu et al. 2016; Wang et al. 2024; Cao et al. 2024). Again, the dependence of $\eta$ on inclination angle and phase are ignored.

\subsection{Emission height as a function of phase}

From equation (\ref{eqn_emission_height}) it can be seen that: (1) The emission height is proportional to $\phi^2$. At the center of the pulse profile, the emission height is lower. And at the edge of the pulse profile, the emission height is higher. (2) The emission height is proportional to $\beta^2$. Pulsars with larger impact angle is expected to have higher emission height. (3) At the same time, pulsars with larger inclination angle $\alpha$ will also tend to have higher emission height. We are not sure how to compare this point with the observations, at present. This is a prediction of the analytical expression. (4) If the emission originates from the fields line near the magnetic axis, i.e. a smaller $\eta$, the pulsar also will have higher emission height. These aspects may be compared with current pulsar observations.

\begin{figure}
  \centering
  \includegraphics[width=0.45\textwidth]{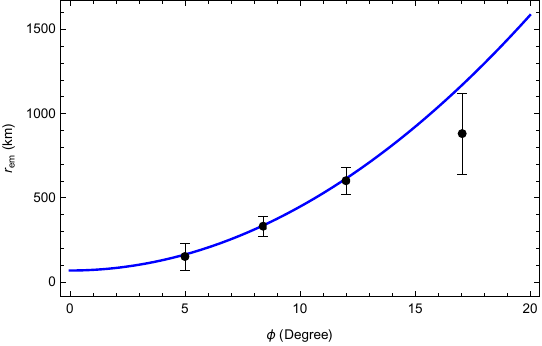}\\
  \caption{Emission height as a function of phase for PSR B0329+54. The black points are the observational data. The blue line is the model calculation (eq.(\ref{eqn_emission_height})).}\label{fig_heightphase}
\end{figure}

Assuming symmetric conal structure, the emission height may be determined by the relativistic phase shift method (Gangadhara \& Gupta 2001). This method is slightly modified later (Dyks et al. 2004). Emission height for different cones is actually emission height at different phases. Our simplified treatment can be compared with the observations of PSR B0329+54 (since it has the maximum number of cones, 4 in total, table 2 in Gangadhara \& Gupta 2001; table 2 in Dyks et al. 2004. These two tables are the main observational data, which are compared with our analytical modeling.). Here only the data of $\nu=325\ \rm Mhz$ are included. The data at other frequencies are similar. Data for other pulsars with less cone (2 or 3 cones) provide less constraint on the model. So we do not include them. The comparison with the observation is shown in figure \ref{fig_heightphase}. The model curve is obtained by using the same $\alpha$ and $\beta$ in Gangadhara \& Gupta (2001). The parameter $\eta$ is set to be $\eta=0.55$. Our value of $\eta$ is consistent with that of Gangadhara \& Gupta (2001) (table 2 there, last column). It is also consistent with the general constraint obtained by Mitra \& Rankin (2002)(section 8 there, where they obtained $\eta$ may lie be between $1/2$ and $2/3$).

A higher emission height at the edge of the pulse profile is also consistent with the observations of pulsars with interpulses (Johnston \& Kramer 2019). By applying the rotating vector model to a group of pulsars with interpulses, the authors try to determine the emission as a distance from the magnetic axis (which can be described by $\theta$ or $\theta_{\rm obs}$, see eq.(\ref{eqn_theta_thetaobs_relation})). A general tendency is seen observationally: As the distance from the magnetic axis increases, the emission height also increases (see figure 5 in Johnston \& Kramer 2019), where a quadratic dependence is assumed. According to our analytical treatment, the emission height is indeed a quadratic function of the magnetic colatitude (see eq.(\ref{eqn_emission_height_thetaobs})).

\subsection{Emission height as a function of impact angle}

\begin{figure}
  \centering
  \includegraphics[width=0.45\textwidth]{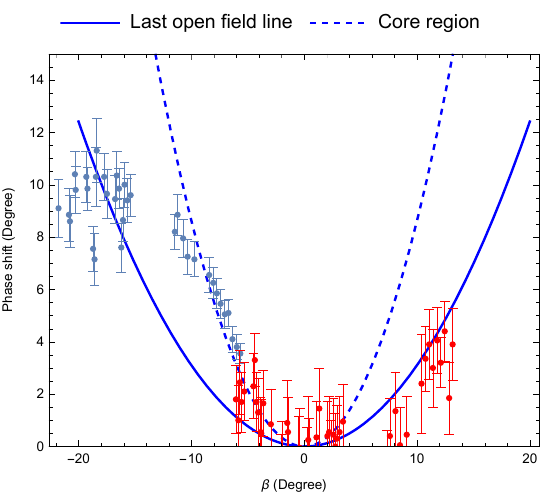}\\
  \caption{Phase shift (or emission height) as a function of the impact angle for PSR J1906+0746. The dark blue data points are for the main pulse, the red data points are for the interpulse. The blue curves are the model calculations (eq.(\ref{eqn_phasebeta})), for both the last open field line and a field line in the core region, respectively.}\label{fig_phasebeta}
\end{figure}

The geodetic precessing pulsar PSR J1906+0746 provides a two-dimensional view of its emission beam (Desvignes et al. 2019). At the same time, by applying the rotating vector model, the geometry and the emission height as a function of impact angle $\beta$ can also be obtained (figure 5 in Desvignes et al. 2019). The emission height is related to the phase shift between polarization position angle curve and pulse profile: $\Delta \phi_s = 4 r/R_{\rm lc}$ (Blaskiewicz et al. 1991). Substituting the equation for the emission height (eq.(\ref{eqn_emission_height})), the model predicted phase shift as a function of phase and impact angle is:
\begin{eqnarray}
  \Delta \phi_s &=& \frac{16}{9} \frac{1}{\eta^2} (\sin^2\alpha \ \phi^2 + \beta^2)\\
  \label{eqn_phasebeta}
  &=& \frac{16}{9} \frac{1}{\eta^2} \beta^2. \quad \quad (\text{for $\phi=0$})
\end{eqnarray}
The comparison with observations is shown in figure \ref{fig_phasebeta}. The observational data points are from Desvignes et al. (2019). In order to compare with the phase shift as a function of impact angle, the phase is set to be $\phi=0$. This is because the method is mainly for the central part of the pulse profile (Blaskiewicz et al. 1991; Dyks et al. 2004; Dyks et al. 2008). The two model curves are: $\eta=1$ for the last open field line, and $\eta=0.6$ for the core region. Here the core field line with $\eta=0.6$ is just a typical calculation. It is not obtained by fitting the observational data.

In Desvignes et al. (2019), figure 5 there, the author cited the theoretical model of Yuen \& Melrose (2014), which is based on Gangadhara (2004). And Desvignes et al. (2019) found some deviations between the observational data and theoretical model: ``close to the beam edge of the main pulse, and large $\beta$ for the interpulse". Our figure 2 shows that if the radio emission at different $\beta$ comes from different field lines (characterized by $\eta$), the observation data and the theoretical curve are consistent each other. From figure \ref{fig_phasebeta}, it may be seen that: (1) when the impact angle $\beta$ is large, the radio emission tends originates from the last open field line. (2) When the impact angle is small, the radio emission may come from field lines in the core region of the polar cap. Geometrically, for a pencil emission beam, it is natural that the emission should come from the boundary field lines (i.e. last open field line) when it is view at a large impact angle.

\subsection{Lorentz factor as a function of phase I: curvature radiation}

When the three spherical coordinate of the emission point is known (especially the height $r$), the geometry of the emission region is known. If an emission mechanism is assumed (e.g. curvature or inverse Compton scattering), then the required electron Lorentz factor can be obtained. On using the emission height as a function of phase (eq.(\ref{eqn_emission_height})), the electron Lorentz factor as a function of phase can be obtained.

\begin{figure}
  \centering
  \includegraphics[width=0.45\textwidth]{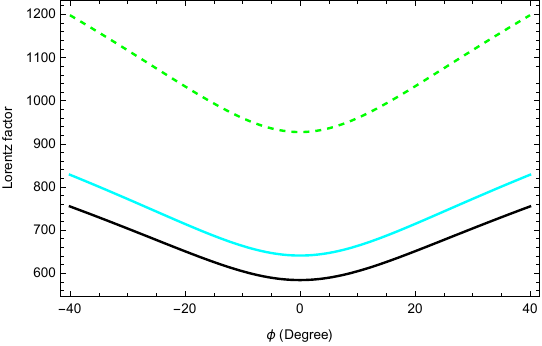}\\
  \caption{Lorentz factor as a function of phase for PSR B0943+10, due to curvature radiation. The three curves from bottom to top are for $\eta=1$ (last open field line), $\eta=0.87$ (annular region) and $\eta=0.5$ (core region), respectively.}\label{fig_Lorentz_factor_phi}
\end{figure}

For a dipole field line, at point $(r,\theta)$, the curvature radius is (Wang et al. 2013, again assuming small $\theta$):
\begin{equation}
  \rho \approx \frac43 R_e \sin\theta = \frac43 \sqrt{r R_e}.
\end{equation}
The corresponding photon frequency due to curvature radiation is (Ruderman \& Sutherland 1975):
\begin{equation}
  \nu = \frac{3 \gamma^3 c}{4\pi \rho}.
\end{equation}
Therefore, the Lorentz factor can be determined to be:
\begin{equation}
  \gamma^3 = \frac{4\pi \rho \nu}{3c} \approx \frac{4\pi \nu}{3c} \frac43 \sqrt{r R_e}.
\end{equation}
Substituting eq.(\ref{eqn_emission_height}) into the above equation, the electron Lorentz factor as a function of phase is:
\begin{flalign}
&\ \gamma = \left( \frac{16}{27} P \nu \right)^{1/3} \eta^{-2/3} (\sin^2\alpha \ \phi^2 + \beta^2)^{1/6} &\\
  \label{eqn_Lorentz_factor_phi}
&\ = 840 \left( \frac{P}{1\rm s} \frac{\nu}{1\rm GHz} \right)^{1/3} \eta^{-2/3} (\sin^2\alpha \ \phi^2 + \beta^2)^{1/6}. &
\end{flalign}
Typical calculations are shown in figure \ref{fig_Lorentz_factor_phi}. The $\alpha$ and $\beta$ of PSR B0943+10 are used during the calculations (Cao et al. 2024, Table 2 there). In this figure, three curves are plotted with $\eta=1, \ 0.87, \ 0.5$. The three values represent, respectively, the last open field line, typical field line in the annular region, and typical field in the core region (Qiao et al.2004; Cao et al.2024). The curves derived from our model are plotted according to our simplified formula, eq.(\ref{eqn_Lorentz_factor_phi}). Figure \ref{fig_Lorentz_factor_phi} can be compared with figure 9, panel (iii) in Cao et al. (2024). There, the Lorentz factor are calculated numerically (see section 2.2 there for discussions). In both Cao et al. (2024) and here, the basic assumption is a dipole field. The radio emission are also supposed to be emitted along the tangent of the field line . In the numerical calculations, a more general parameter space may be explored. However, the radio emission beam radius and the impact angle, etc, are often small quantities for most pulsars. Therefore, our analytical treatment is consistent with the numerical calculations.

\begin{figure}
  \centering
  \includegraphics[width=0.45\textwidth]{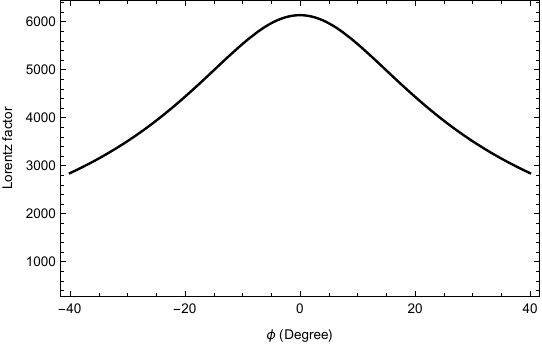}\\
  \caption{Lorentz factor as a function of phase, for the inverse Compton scattering case.}\label{fig_Lorentz_factor_phi_ICS}
\end{figure}

\subsection{Lorentz factor as a function of phase II: inverse Compton scattering}

The curvature radiation is only one of the coherent radio emission mechanisms by particle bunches. For relativistic particles, in the presence of various kinds of low frequency waves, the inverse Compton scattering may be an alternative coherent radio emission mechanism (Qiao \& Lin 1998; Qiao et al. 2001; Xu et al. 2000). Recently the inverse Compton scattering is also applied to the case of fast radio bursts (Zhang 2022 and references therein for a survey of previous works).

For a low energy seed photon with frequency $\nu_0$, the final scattered photon will have a frequency (Qiao \& Lin 1998; Zhang 2022):
\begin{equation}
  \nu \approx \gamma^2 \nu_0 (1- \cos\theta_i),
\end{equation}
where $\theta_i$ is the angle between incoming photons wave vector and electron momentum, and the dimensionless electron velocity is taken as $1$. Assuming that the radio emission height is significantly larger than the neutron star radius, the incoming photon may be viewed as propagating along the radial direction. While the relativistic electrons move along the magnetic field lines, then $\theta_i$ will be:
\begin{equation}
  \theta_i =\theta_{\rm obs} - \theta = \frac12 \theta =\frac13 \theta_{\rm obs}.
\end{equation}
Then the final scattered photon frequency is:
\begin{equation}
  \nu \approx \gamma^2 \nu_0 \frac12 \theta_i^2 = \gamma^2 \nu_0 \frac{1}{18} (\sin^2\alpha \ \phi^2 + \beta^2).
\end{equation}
The Lorentz factor as a function of phase in the inverse Compton scattering scenario is:
\begin{flalign}
 &\  \gamma = \left( 18 \frac{\nu}{\nu_0} \right)^{1/2} (\sin^2\alpha \ \phi^2 + \beta^2)^{-1/2} \qquad\qquad\qquad & \\
 &\ = 1340 \left( \frac{\nu}{10^9 \ \rm Hz} \frac{10^4 \ \rm Hz}{\nu_0} \right)^{1/2} (\sin^2\alpha \ \phi^2 + \beta^2)^{-1/2}. &
\end{flalign}
A typical calculation is shown in figure \ref{fig_Lorentz_factor_phi_ICS}, using the same $\alpha$ and $\beta$ of PSR B0943+10. As can be seen from figure \ref{fig_Lorentz_factor_phi_ICS}, the required Lorentz factor is typically higher than the curvature radiation case. This point is also noted in previous works (in pulsars, Qiao et al. 2001; in fast radio burst, a higher parallel electric field is required, Zhang 2022).
A second difference is that the Lorentz factor is higher at the center of the pulse profile. This is opposite to that of the curvature radiation (compare figure \ref{fig_Lorentz_factor_phi} and \ref{fig_Lorentz_factor_phi_ICS}). At present, we do not know how to test this point observationally.
We guess that: (1) since the inverse Compton scattering requires a higher Lorentz factor at the center, it may result in a steeper spectra for the core component (Lyne \& Manchester 1988). (2) The required Lorentz factor at the edge is lower, this may help to make a wider emission beam, e.g. in the fan beam scenario (Wang et al. 2014). These points will require some knowledge about the spatial and energy distribution of the relativistic particles. While, at present, we mainly focus on the geometry of the emission beam.
Here the field line parameter $\eta$ does not enter into the equations. The main input parameter is $\theta_i$. However, we speculate that the electrons may move mainly along the last open field line.

\section{Discussion and conclusions}

In the above, we try to provide an analytical treatment of pulsar radio emission height as a function of phase based on a set of simplifying assumptions. In order to fulfill the ``analytical'' requirement, a dipole field line geometry is assumed. At the same time, the three quantities $\theta,\ \phi, \ \beta$ are assumed to be small quantities. If only two or one of these quantities can be assumed as small quantity, then similar analysis may also be obtained. For example, in the one pole scenario for interpulses (Lyne \& Manchester 1988; Sun et al. 2025), the phase $\phi$ can not be assumed as small quantity, but $\theta$ and $\beta$ can still be assumed as small quantities. Then the only modification is to replace eq.(\ref{eqn_thetaobs}) with:
\begin{equation}
\theta_{\rm obs}^2 = \sin\alpha \sin(\alpha + \beta) 4 \sin^2\frac{\phi}{2} + \beta^2.
\end{equation}
The rest will be the same.
In the two pole scenario for the interpulse (Wang et al. 2024; Sun et al. 2025), if only the main pulse or interpulse is concerned at one time (neglecting the bridge emission), then the above treatment is still valid (e.g. eq.(\ref{eqn_emission_height})). The only requirement is that: the corresponding $\alpha$ and $\beta$ for the main pulse or interpulse should be employed.

The final results of the analytical treatment can be found in a collection of previous works. The merit of the analytical treatment is its simplicity: (1) From the analytical expression, it can be seen that the emission height as a function of phase or impact angle belong to the same physics. (2) It can applied directly to related physics, e.g., the Lorentz factor as a function of phase. (3) The difference between the curvature radiation and inverse Compton scattering can be seen directly. Of course, there are assumptions (dipole field) and approximations (small beam radius and impact angle etc), as has been discussed above.

Our analytical treatment (on emission height and Lorentz factor) can be compared or used to analyze the observations directly. After this is done, if more accurate treatment are needed, then more complicated field line geometry (Perera et al. 2014; Huang et al. 2016) and more accurate numerical calculations can be employed (e.g., Zhang et al. 2007; Lee et al. 2009; Cao et al. 2024 and references therein). A simplified treatment of the theoretical model may be more urgent in the FAST and SKA era, when more high quality observations of more source are available (Han et al. 2021).

Previously, which field line does the radio emission come from is not known exactly. This is reflected in the parameter $\eta$ in eq.(\ref{eqn_Re}). The last open field line may be not the only choice (Gangadhara \& Gupta 2001; Mitra \& Rankin 2002; Wang et al. 2006; Zhang et al. 2007; Wang et al. 2013; Wang et al. 2024; Cao et al. 2024). Furthermore, whether the field line lies in the annular region or core region of the polar cap may result in different physics of particle acceleration (Qiao et al. 2004; Wang et al. 2024; Cao et al. 2024). The result from figure 2 may imply that pulsars with large impact angle may radiate from the last open field line. While pulsars with small impact angle may radiates from regions inside the polar cap. Of course, the exact value of $\beta$ may not be important. The ratio of $\beta/\rho_{\rm beam}$ may be more relevant, where $\rho_{\rm beam}$ is the beam radius of the radio emission. At present only a tendency is seen in figure \ref{fig_phasebeta}. We are not sure about further details.

In summary, under some approximations, we have obtained an analytical treatment for the pulsar radio emission height as a function of phase (eq.(\ref{eqn_emission_height})). This can be compared with (1) the observations of emission height as a function of phase, (2) emission height (or in terms of phase shift) as a function of the impact angle, (3) Lorentz factor as a function of phase. Our simplified treatment are consistent with the observations and numerical calculations.

\section*{acknowledgments}
The authors would like to thank Dr. Gregory Desvignes very much, for providing the observational data which are used in figure \ref{fig_phasebeta}, and the referee for helpful suggestions which improved the quality of this paper. This work is supported by National SKA Program of China (No. 2020SKA0120300) and NSFC (12133004).




\label{lastpage}

\end{document}